\documentstyle[aps,eqsecnum,preprint,prd,epsfig,epsf]{revtex}
\begin{document}
\addtolength{\jot}{10pt}
\tighten
\newcommand{\spur}[1]{\not\! #1 \,}
\newcommand{\be}{\begin{equation}}
\newcommand{\ee}{\end{equation}}
\newcommand{\bea}{\begin{eqnarray}}
\newcommand{\eea}{\end{eqnarray}}

\draft
\preprint{\vbox{\hbox{BARI-TH/99-351 \hfill} 
                \hbox{UGVA-DPT 1999/09-1051 \hfill}
                \hbox{August 1999\hfill} }}
\vskip 1cm
\title{\bf Using  Heavy Quark Spin Symmetry\\
in Semileptonic $B_c$ Decays}
\author{Pietro Colangelo$^a$ and 
Fulvia De Fazio\footnote{``Fondazione Angelo Della Riccia'' Fellow}$^b$\\}

\vskip 1.0cm

\address{
$^a$ Istituto Nazionale di Fisica Nucleare, Sezione di Bari, Italy\\
$^b$ D\'epartement de Physique Th\'eorique, Universit\'e de Gen\`eve, 
Switzerland
\\}

\maketitle
\vskip 0.5cm
\begin{abstract}
The form factors parameterizing the $B_c$ semileptonic 
matrix elements  can be related to
a few invariant functions if the decoupling of the spin of the heavy quarks
in $B_c$ and in the mesons produced in the semileptonic decays is exploited.
We compute the form factors as overlap integral of the meson wave-functions 
obtained using a QCD relativistic potential model, and give 
predictions for semileptonic and non-leptonic $B_c$ decay modes. 
We also discuss possible experimental tests of the heavy quark spin
symmetry in $B_c$ decays.
\end{abstract}
\vspace*{1cm}
\clearpage

\section{Introduction}\label{s:intr}

The discovery of the $B_c^+$ meson by the CDF Collaboration at
the Fermilab Tevatron \cite{cdf} opens for interesting
investigations concerning  the structure of
strong and weak interactions in the
quarkonium-like $\bar b c$ hadronic system. The studies will be
further
developed at the hadronic machines currently under construction, such as
the LHC accelerator at CERN, where a copious production of $B_c$ meson
and of its radial and orbital excitations is
expected \cite{bcprod,revbc}; 
at these experimental facilities, together with the
measurement of the mass of the  particles belonging to  the 
$\bar b c$ $(b \bar c)$ family, it will be possible to observe the decay
chains reaching the $^1S_0$ ground state, the $B_c$, which decays weakly. 

A peculiarity of the $B_c$ decays,
with respect to the decays of the 
$B_{u,d}$  and $B_s$ mesons, is that both the quarks  are involved in the
weak
decay process with analogous probability. The weak decays of the charm
quark, whose mass is
lighter than the $b$ quark mass, are   mainly governed by the
CKM matrix element $V_{cs}$ which is larger than $V_{cb}$ mainly
controlling the $b$ quark transitions; the result is that both the quark
decay processes contribute on a comparable footing to the $B_c$  decay
width. Another peculiar aspect is that  the $\bar b c$ annihilation amplitude,
proportional to $V_{cb}$, is enhanced with respect to the 
analogous amplitude describing the $B^+$ annihilation mode.

The above considerations have inspired several theoretical analyses 
\cite{lus,quigg,paver,beneke,ani} aimed at
predicting the  $B_c$ lifetime.
Namely, a QCD analysis \cite{beneke}, based on the OPE expansion in
the inverse
mass of the heavy quarks and on the assumption of quark-hadron duality,
provides for $\tau_{B_c}$ a prediction in agreement 
(at least within the current experimental accuracy)
with the CDF measurement:
$\tau(B_c) = 0.46^{+0.18}_{-0.16} \; ({\rm stat}) 
\pm 0.03 \; ({\rm syst}) 10^{-12} {\rm s}$ \cite{cdf}.
The agreement supports 
the overall picture of the inclusive $B_c$ decays.

The calculation of
the $B_c$ exclusive decay modes can be carried out either using
QCD-based methods,
such as lattice QCD or QCD sum rules, or adopting some constituent quark
model.
So far, lattice QCD has only been employed to calculate the 
$B_c$ purely leptonic width \cite{woloshyn}.
As for QCD sum rules \cite{svz}, 
the $B_c$ leptonic constant, as well as the matrix
elements relevant for the semileptonic decays, were computed in 
refs.\cite{oldbc,paver,bagan}. These analyses  identified  a difficulty
in correctly considering the
 Coulomb pole contribution in the three-point functions needed for
the calculation of the semileptonic matrix elements.
Attempts aimed at taking this correction
into account are described in
\cite{kiselev}; however,
the problem of including the contribution of the Coulomb pole 
for all the values of
the squared momentum transfer $t$ to the lepton pair has not been solved,
yet. Extending to all values of $t$ the expression of the Coulomb contribution
valid at $t_{max}$ only allows  to conclude that it
represents a large correction  to the lowest order quark
spectral functions. 

It is worth looking at the outcome of constituent quark models which,
although less established on the QCD theoretical ground, can
nevertheless provide us with significant information to be
compared to the experimental results. 

The models in refs.\cite{bsw,isgw} have been used in the past \cite{lus,du}
to estimate
the semileptonic $B_c$ decay rates. More recently, different versions of the
constituent
quark model have been used to analyze the decays
induced both by the $b \to c(u)$ and $c \to s(d)$ transitions
\cite{chang,liu}. It is noticeable  that the calculations
can be put on a firmer theoretical ground if some dynamical features of
the $B_c$ decays are taken into account. Such features are mainly
related to the decoupling of the spin of the heavy quarks of the $B_c$
meson, as well as of
the meson produced in the semileptonic decays, i.e.
mesons belonging to the $\bar c c$ family ($\eta_c, J/\psi$, etc.) and
mesons containing a single heavy quark
($B_s^{(*)}$, $B_d^{(*)}$, $D^{(*)}$).
The decoupling occurs in the heavy quark  limit
($m_b, m_c \gg \Lambda_{QCD}$), and produces
a symmetry, the heavy quark spin symmetry, allowing to relate the form
factors governing the $B_c$ decays into a $0^-$ and $1^-$ final meson
to a few invariant functions
\cite{jenkins}. The main consequence is that the number of form factors 
parameterizing the  matrix elements
is reduced, and the description of the semileptonic transitions is greatly
simplified. 

However, at odds of the heavy quark flavour symmetry, holding for
heavy-light mesons, spin symmetry does not fix 
the normalization of the form factors at any point of the phase space.
The normalization, as well as the functional dependence near the zero-recoil
point, must be
computed by some nonperturbative approach.

So far, the ``universal'' form factors of semileptonic $B_c$ decays have been 
estimated using nonrelativistic meson wave-functions \cite{jenkins} and
employing the ISGW   model at the zero-recoil point 
\cite{sanchiz}. An analysis in the framework of a different quark model is 
described in \cite{chang}.

In this paper we present a calculation based on a
constituent quark model which has been used to describe several
aspects of  the heavy meson  phenomenology \cite{pietroni}. The
peculiar features of the model are related to the interquark potential,
which follows general QCD properties, such as 
scalar flavour-independent
confinement at large distances, and asymptotically free  QCD  coulombic
behaviour at short distances. Moreover, the use of the relativistic form
of the quark kinematics allows to describe 
heavy-light  as well as   heavy-heavy mesons,  and to account for
deviations from the nonrelativistic limit.
As a result,  the $B_c$ form factors can be written as overlap integrals of
meson wave-functions, obtained by solving the wave equation defining
the model.  As discussed in the following, 
the representation as overlap integral of meson wave-functions allows to 
predict, in the heavy quark limit, the normalization of the invariant 
functions at the zero-recoil point and to obtain, for example, 
the suppression factor between the form factors of the $B_c$ transitions into 
heavy-light mesons with respect to the corresponding functions governing the
decays
$B_c\to \eta_c \ell \nu$ and $B_c\to J/\psi \ell \nu$.

The calculation of the overlap integrals and of the $B_c$ semileptonic
form factors
is presented in Sec. III, after having reviewed in Sec. II the
consequences of the heavy quark spin symmetry in $B_c$ decays. In Sec.   
IV, using the obtained invariant functions,
 we analyze the semileptonic decay modes, and in Sec. V,
assuming the factorization ansatz, we estimate
several non-leptonic $B_c$ decay rates. Sec. VI is devoted to the
conclusions.

\section{Heavy Quark Spin Symmetry}\label{s:symmetry}

Heavy quark spin symmetry  amounts to
assume the decoupling between the spin of the heavy quarks
in the $B_c$ meson, since the
$\bar b c$ spin-spin interaction vanishes in the infinite heavy quark
mass limit, as well as  the vanishing of the heavy quark-gluon 
vertex.
This symmetry has been invoked in \cite{jenkins} to work out relations among
the semileptonic 
matrix elements between $B_c$ and other heavy mesons
(both
heavy-heavy and heavy-light). The main difference with respect to the most
well known case of the heavy-light systems is that in the latter case
one can exploit heavy quark flavour symmetry, which also holds in the
heavy quark limit and allows to relate $B$ to $D$ form factors. 

In order to apply spin symmetry to $B_c$ decays one should 
distinguish decays due to charm  transitions from $b$ quark
transitions.
To the first category belong processes such as $B_c \to (B_s, B_s^*)\ell \nu$ 
and 
 $B_c \to (B_d, B^*_d) \ell \nu $,                      
induced at the quark level by the
transitions $c\to s$ and $d$, respectively.
Since $m_c \ll m_b$, the energy released in such decays to the final hadronic 
system  is much less than
$m_b$, and therefore the $b$ quark remains almost unaffected. As a
consequence, the final $B_a$ meson ($a$ is a light $SU(3)_F$ index) keeps 
the same $B_c$ four-velocity $v$, apart
from a small residual momentum $q$.
The initial and final meson momenta can then be written as:
 $p_{B_c}=M_{B_c}v$ and
$p_{B_a}=M_{B_a}v+q$, with $v \cdot q={\cal O}\big({1 \over m_Q} \big)$.
The relation between the residual momentum
$q$ and the momentum $k$ transferred to the lepton pair is
\be
k^\mu=p_{B_c}^\mu-p_{B_a}^\mu=(M_{B_c}-M_{B_a})v^\mu-q^\mu \;. 
\label{k}
\ee
In this kinematic situation, exploiting the decoupling of the
spin of the heavy quarks in the mesons, several
relations can be worked out among
the semileptonic $B_c$ form factors. A straightforward way to derive such
relations is to use the trace formalism \cite{trace,neu} 
\footnote{For a discussion of the heavy quark formalism applied to the 
quarkonium system see ref. \cite{cas} and references therein.}.
This has been done in ref. \cite{jenkins}, and  we repeat here the derivation 
for the sake of completeness. 
 
One introduces  a $4\times 4$ matrix  $H^{c {\bar b}}$
describing the doublet $(B_c,B_c^*)$ of
$c {\bar b}$ mesons of four-velocity $v$ \cite{jenkins}:
\be
H^{c{\bar b}}={(1+\spur v) \over 2}[B_c^{*\mu} \gamma_\mu - B_c
\gamma_5]
{(1-\spur v) \over 2}\;  \;\; , \label{hbc}
\ee
where $B_c^{*\mu}$ and $B_c$ annihilate a vector $B^*_c$ and a 
pseudoscalar $B_c$ meson of four-velocity $v$.
Under spin rotations of the heavy quarks, $H^{c {\bar b}}$
transforms as
$H^{c \bar b} \to S_c H^{c \bar b} S^\dagger_{\bar b}$.

On the other hand, for heavy-light  $B_a$ and $B_a^*$ mesons,
the analogous $4 \times 4$ matrix describing the $(B_a,B_a^*)$ spin multiplet
reads:
\be
H_a={(1+\spur v) \over 2}[B_a^{*\mu} \gamma_\mu - B_a
\gamma_5] \;\;\; ;
\label{ha}
\ee
all the fields in (\ref{hbc}),(\ref{ha})
contain a factor $\sqrt{M_{B_{c,a}}}$  and
have therefore dimension $3/2$.

Applying  the trace formalism, one gets
that the hadronic matrix elements relative to the decays 
$B_c \to B_a^{(*)} \ell \nu$ have the following general form, compatible with
heavy quark spin symmetry:
\be
<B_a^{(*)},v,q|\bar q_a \Gamma c| B_c, v> = - \sqrt{M_{B_c} M_{B_a}} 
Tr [ {\bar H_a} \Omega \Gamma H^{c \bar b}]
\ee
where $\Omega$ is the most general Dirac matrix proportional to the 
four-velocity $v$ and to the residual momentum $q$. The calculation using 
(\ref{hbc}),(\ref{ha}) shows that the various matrix elements reduce to:
\bea
<B_a,v,q|V_\mu|B_c,v>&=& \sqrt{2M_{B_c}2M_{B_a}}[\Omega_1^a~ v_\mu +a_0~
\Omega_2^a~ q_\mu] \; , 
\nonumber \\
<B_a^*,v,q|V_\mu|B_c,v>&=& -i \; \sqrt{2M_{B_c}2M_{B^*_a}}~a_0~
\Omega_2^a~ \epsilon_{\mu \nu \alpha \beta} \epsilon^{*\nu} q^\alpha v^\beta
\;, \label{ba} \\
<B_a^*,v,q|A_\mu|B_c,v>&=& \sqrt{2M_{B_c}2M_{B^*_a}}[\Omega_1^a
~\epsilon^*_\mu
+a_0~
\Omega_2^a~ \epsilon^* \cdot q ~v_\mu] \; ,
\nonumber 
\eea
\noindent where $V_\mu$ and $A_\mu$ represent the weak
flavour-changing ($c \to s,d$)
vector and axial current, respectively, and $\epsilon$ is the $B^*_a$ 
polarization vector. 
Therefore, as shown by eq.(\ref{ba}),
the six form factors parameterizing the $B_c$ 
into $B_a$ and $B_a^*$ matrix elements 
can be expressed in terms of  two invariant
functions,
$\Omega_1^a$ and $\Omega_2^a$. The main difference with respect to the
spin-flavour symmetry, holding in heavy-light mesons, is that the
normalization of the form factors is not predicted at any point of the
kinematic  range and, in particular, it is not fixed at the non-recoil point 
$q=0$.

Actually, the form factors
$\Omega_2^a$ give rise to
terms proportional to the lepton mass in the calculation of the semileptonic 
rates. Moreover,
$\Omega_2^a$ do not contribute at zero-recoil. 
The scale parameter $a_0$ is related to the size of the
$B_c$ meson, it can be assumed as 
proportional  to the $B_c$ Bohr radius and represents the typical range of 
variation of the form factors \cite{jenkins}.

The relations (\ref{ba}) are valid near the zero-recoil point, where both
$B_c$ and the meson produced in the decay are nearly at rest. In the case
of the transitions $B_c \to B_s^{(*)}, B_d^{(*)}$ the physical phase space
is quite narrow (the maximum momentum transfer $t$ to the lepton pair
 is $t_{max}\simeq 1 $ GeV$^2$) and therefore one can assume that
eqs.(\ref{ba}) completely determine the semileptonic matrix elements
(modulo a set of corrections mentioned below). The situation is different
for processes induced, at the quark level, by the $b-$quark transitions.
Let us consider the decays
$B_c \to (D,D^*)\ell \nu$,
 induced by the $b \to u$ transition. In this case,
the energy released to the final meson is small
only near the zero-recoil point, where
$q^2 \ll m_c^2$. At such kinematic point one can repeat the considerations 
for the transition $B_c \to B_s\ell \nu$, obtaining the relations:
\bea
<D,v,q|V_\mu|B_c,v>&=& \sqrt{2M_{B_c}2M_D}[\Sigma_1~ v_\mu +a_0
~\Sigma_2~ q_\mu] \; , 
\nonumber \\
<D^*,v,q|V_\mu|B_c,v>&=& -i \; \sqrt{2M_{B_c}2M_{D^*}}~a_0~
\Sigma_2 ~\epsilon_{\mu \nu \alpha \beta} \epsilon^{*\nu} q^\alpha v^\beta
\;, \label{d} \\
<D^*,v,q|A_\mu|B_c,v>&=& \sqrt{2M_{B_c}2M_{D^*}}[\Sigma_1~ \epsilon^*_\mu
+a_0~
\Sigma_2~ \epsilon^* \cdot q ~v_\mu] \; .
\nonumber
\eea
Far from the non-recoil point, the light recoiling quark keeps a large 
momentum, and therefore terms of the order of $q\over m_c$ cannot be 
neglected in the effective theory leading to (\ref{d}). 

Finally, we consider $B_c$ decays into quarkonium states, such as
$\eta_c$ and $J/\psi$. The
spin decoupling  of both the beauty and charm quark  allows now to
relate the six form factors to a single one:
\bea
<\eta_c,v,q|V_\mu|B_c,v>&=& \sqrt{2M_{B_c}2M_{\eta_c}} ~\Delta ~v_\mu
\nonumber \\
<J/\psi,v,q|A_\mu|B_c,v>&=& \sqrt{2M_{B_c}2M_{J/\psi}} ~\Delta 
~\epsilon^*_\mu \;\;\; . \label{qqbar}
\eea

Also in this case eqs.(\ref{qqbar}) are only valid near 
the zero-recoil point. Nevertheless,  
in the following we use them, as well as eqs. (\ref{d}),
for all physical values of the momentum transfer $t$, 
in order to compute semileptonic and non-leptonic $B_c$ decay rates.
This is  admittedly a strong assumption, 
and the related uncertainty must be added 
to the uncertainties coming from finite mass and QCD 
corrections that in principle relate
the invariant functions to the physical semileptonic
matrix elements \cite{jenkins}.
However, assuming eqs.(\ref{qqbar}) and (\ref{d}) in the whole kinematic 
range, a number of predictions can be collected; 
the experimental results will then provide us with indications on the 
numerical importance of the corrections.

\section{$B_c$ form factors from a constituent quark model}\label{s:model}

In this section we  compute the
form factors $\Delta$, $\Omega_1^a$ and $\Sigma_1$ by using
a relativistic potential model which allows to account for two
QCD effects. The first one is
confinement, which produces a suppression, at large distances, of the
meson wave-functions, 
due to the linearly increasing interquark potential. The second effect
is represented by the deviation of the quark dynamics from the 
nonrelativistic limit. By taking  such two effects into account,
we are able to compute the form factor $\Delta$ in (\ref{qqbar}) 
as an overlap integral of $B_c$ and $J/\psi$ wave-functions. 
Moreover, we can
apply the formalism to the transitions 
$B_c \to B_s^{(*)},  B_d^{(*)}$ and $D_d^{(*)}$ at the 
non-recoil point, and then  extrapolate the result to the whole kinematic 
region spanned by the various semileptonic transitions.

Let us consider  $\Delta$ in (\ref{qqbar}).
In order to compute it,  we consider the costituent quark model
studied in \cite{pietroni},  whose  essential features 
can be easily summarized.
First,  we write down an expression for
the $B_c^+$ meson state, in the $B_c^+$ rest frame, in terms of quark and
antiquark creation operators, and of a meson wave-function:
\begin{equation}
|B_c^+>=i {\delta_{\alpha \beta} \over \sqrt{3}}
{\delta_{rs} \over \sqrt{2}}
\int d \vec{k} \; \psi_{B_c} (\vec{k})\;
b^{\dag}(-\vec{k}, r, \alpha) \;
c^{\dag}(\vec{k}, s, \beta)|0>  \label{b}
\end{equation}
where $\alpha$ and $\beta$ are colour indices, $r$ and $s$ spin indices.
The operator $b^{\dag}$ creates an anti-$b$ quark  with momentum
$-{\vec k} $, while $c^{\dag}$ creates a charm quark
with momentum $\vec{k}$. A similar expression holds for the $\eta_c$ 
($\bar c c$) state,
as well as for  vector $1^-$ states, as described in \cite{pietroni}.
In the meson state,  as written
in (\ref{b}),  the contribution
of other Fock states such as, e.g.,  states containing one or more
gluons, is neglected.

The wave-function  $\psi_{B_c} (\vec k)$ describes 
the momentum distribution of the quarks in the meson. It 
is obtained by solving the wave equation 
\begin{equation}
\Big\{ \sqrt{\vec{k}^2 + m_b^2} + \sqrt{\vec{k}^2 + m_c^2}-M_{B_c} \Big\}  
\psi_{B_c} ( \vec{k})
+ \int d \vec {k^{\prime}} \; V(\vec{k}, \vec{k^{\prime}}) \;
  \psi_{B_c} (\vec{k^{\prime}})=0 
\label{7} 
\end{equation}
stemming from the quark-antiquark
Bethe-Salpeter equation, in the approximation of an
istantaneous interaction represented by the  potential $V$. 
Eq.(\ref{7}) partially takes into account the relativistic
behaviour of the quarks in the kinetic term; 
$m_c$ and $m_b$ represent the mass of the constituent charm and beauty quark, 
and $M_{B_c}$ the mass of the bound state.

The QCD interaction is described assuming a static interquark
potential having the form, in the coordinate space \cite{richardson}:
\begin{equation}
V(r)={8 \pi \over 33 -2 n_f} \Lambda \Big[ \Lambda r-{f(\Lambda r) \over
\Lambda r} \Big] \;\;\; , \label{pot}
\end{equation}
\noindent with $\Lambda$  a scale parameter,
$n_f$  the number of active flavours, and the function $f(t)$  given
by
\begin{equation}
f(t)={4 \over \pi} \int_0^\infty dq {sin(qt) \over q} \Big[ {1 \over
ln(1+q^2)}
- {1 \over q^2} \Big] \; . \label{f}
\end{equation}
\noindent
The interest for this form of the potential is that it continuously 
interpolates the linearly confining behaviour at large distances with the
QCD coulombic behaviour at short distances, where
 the logarithmic reduction of the strong coupling constant,  due to the 
asymptotic freedom property of QCD, is implemented. A further smoothing of the
potential at short distances is adopted, 
according to quark-hadron duality arguments \cite{pietroni}.

The wave equation (\ref{7}), together with the form (\ref{pot}) of the 
potential and (\ref{b}) of the meson state, completely determines the model,
which  has been  extensively studied to describe static as well as
dynamic properties of mesons containing heavy quarks
\cite{tedesco,defazio,fdf}. Notice that the
spin interaction effects are neglected since,
in the case of heavy mesons, the chromomagnetic coupling is of the order
of the inverse heavy quark masses. Therefore, 
 both the pseudoscalar and the vector mesons, being degenerate in mass, 
are described by the same  wave-function. 

An equation for the form factor $\Delta(\vec q =0)$ in  (\ref{qqbar})
can be obtained expressing 
the $b \to c$ flavour-changing weak currents in terms of quark 
and antiquark operators; for the vector current,  the expression is
\begin{eqnarray}
V^\mu={\delta_{\alpha \beta}\over (2 \pi)^3} \int d \vec q d {\vec q}^\prime
 \Big[{m_b m_c \over E_b(\vec q) E_c({\vec q}^\prime)} \Big]^{1\over 2} 
:&[&\bar u_b (\vec q, r) b_b^\dagger(\vec q, r, \alpha) +
\bar v_b (\vec q, r) d_b(\vec q, r, \alpha)] \gamma^\mu \nonumber \\
&[&u_c ({\vec q}^\prime, s) b_c({\vec q}^\prime, s, \beta) +
\bar v_c ({\vec q}^\prime, s) d_c^\dagger({\vec q}^\prime, s, \beta)]:
\end{eqnarray}
($E_q(\vec k)=\sqrt{k^2+m_q^2}$, $k=|\vec k|$);
an analogous expression describes  the axial current. Then, writing down
the matrix elements (\ref{qqbar}) and
applying canonical anticommutation relations \cite{pietroni,tedesco}, we 
obtain:
\begin{equation}
\Delta(\vec q=0) = {1 \over 2 \sqrt { 2 M_{B_c} 2 M_{\eta_c}} }
\int_0^\infty dk { u_{B_c} (k) u_{\eta_c} (k) \over \sqrt{E_b E_c} }
{ (E_b+ m_b)(E_c+ m_c)-k^2 \over [(E_b+ m_b)(E_c+ m_c)]^{1/2}} \;\;\; ,
\label{delta}\end{equation}
where the reduced wave-functions $u_M(k)$ are related to the $L=0$ 
wave-functions $\psi_M$ according to
\begin{equation}
u_{M}(k)={k \; \psi_{M}(|\vec k|) \over \sqrt{2} \pi } \hskip 15 pt .
\label{eq : 18} 
\end{equation}
The covariant normalization is adopted: 
$\int_0^\infty dk |u_{M}(k)|^2=2 M_{M}$.

The wave-functions $u_{B_c}$ and $u_{\eta_c}$ can be 
obtained by solving eq.(\ref{7})  
 by numerical methods, choosing the values of 
the masses $m_c$ and $m_b$ of the constituent quarks,
together with  the scale parameter $\Lambda$, in such a
way that the charmonium and  bottomonium spectra 
are reproduced: $m_b= 4.89$ 
GeV and  $m_c=1.452$ GeV, with $\Lambda=397$ MeV \cite{pietroni}.
A fit of the heavy-light meson masses also fixes the
values of the constituent light-quark masses: $m_u=m_d=38$  MeV and
$m_s=115$ MeV
\cite{pietroni}. It is worth observing that,
for the $\bar b  c$ system, all the input parameters needed in 
(\ref{7}) are fixed from the analysis of other channels, and the predictions 
do not depend on new external quantities. 

The numerical solution of (\ref{7})
produces  the spectrum of the $\bar b c$ bound states; 
the predicted mass and the leptonic constant of the 
first  $S-$wave resonance are \cite{fdf}:
$M_{B_c}=6.28$ GeV (the value  we use in our analysis) 
and $f_{B_c}=432$ MeV, 
in agreement with  other theoretical 
determinations based on constituent quark models
\cite{qmodels},  QCD sum rules ($M_{B_c}=6.35$ GeV \cite{paver})
and lattice QCD ($M_{B_c}=6.388\pm 9\pm 98\pm 15$ GeV \cite{shanahan}).
Within the errors, the 
$B_c$ mass   agrees with the CDF result:
$M_{B_c}= 6.40 \pm 0.39\; ({\rm stat})\; \pm  0.13 \; ({\rm syst})$ GeV
\cite{cdf}.

The obtained $B_c$ wave-function $u_{B_c}(k)$ is depicted in fig.1. In the 
same figure we plot the wave-functions of the other mesons involved in $B_c$ 
semileptonic decays: $B_s$ and $B_d$, the $\bar c c$ states $\eta_c$ and 
$J/\psi$ together with the first radial excitation  $\eta_c^\prime$ and 
$\psi(2S)$, and the  $D$ meson.

Let us come back to eq.(\ref{delta}) which provides the form factor $\Delta$. 
For quark masses larger than the typical relative quark-antiquark momentum
$k$, eq.(\ref{delta}) becomes:
\begin{eqnarray}
\Delta(\vec q = 0) &=& {1 \over (2 \pi)^3 } 
{1 \over \sqrt { 2 M_{B_c} 2 M_{\eta_c}} }
\int d \vec k \;\; \psi_{B_c} (\vec k) \;\;\psi_{\eta_c}^* (\vec k)
\nonumber \\
&=& {1 \over \sqrt  {2 M_{B_c} 2 M_{\eta_c}}}  
\int d \vec x \;\; \Psi_{B_c} (\vec x) \;\; \Psi_{\eta_c}^* (\vec x) 
\;\;\;\: ,
\label{delta1} 
\end{eqnarray}
where   $\Psi_M(\vec x)$ is defined as
\begin{equation}
\Psi_M(\vec x) = {1 \over (2 \pi)^3 } \int d \vec k \;\;
e^{i \vec k \cdot\vec x} \;\;
\psi_M(\vec k) \;.
\end{equation}
Eq. (\ref{delta1}) shows that
 the form factor $\Delta$, at the zero-recoil point, 
is simply given by the overlap integral 
of the $B_c$ and $\eta_c$ wave-functions
in the coordinate space. 
This result has already been obtained in \cite{jenkins}, 
as it is typical of the calculation of form factors by  
quark models \cite{tedesco,francesi}. The interest in eq.(\ref{delta1})
is that no factors appear in the integral other than the wave functions; this
implies that, in the limit where the $B_c$ and $\eta_c$ wave-functions are 
equal (modulo the normalization condition),  the form factor $\Delta$ is 1.
Although
such an overlap is not constrained by symmetry arguments, as  in the
case of the flavour symmetry in heavy-light mesons, from 
eq.(\ref{delta1}) it turns out that the deviation from unity of the 
invariant function  at the zero-recoil point 
is due to the actual shapes of the meson wave-functions.
In our specific case, as reported in Table \ref{t:tab0}, the deviation from
unity is a  $5 \%$ effect.

The calculation of $\Delta$ near the zero-recoil point,
for a small momentum $\vec q$,  can be
performed by 
modifying eq.(\ref{delta1}), as discussed in \cite{jenkins}:
\begin{equation}
\Delta(\vec q) = {1 \over \sqrt {2 M_{B_c} 2 M_{\eta_c}}} 
\int d \vec x \; e^{i \vec q \cdot \vec x /2} \; 
\Psi_{B_c} (\vec x) \; \Psi^*_{\eta_c} (\vec x) \;\;\; ,
\label{delta2} \end{equation}
and using the relation (valid near the zero-recoil point)
$y= {p_{B_c} p_{\eta_c} \over M_{B_c} M_{\eta_c}}=
\sqrt{1+ \vec q^2/ M_{\eta_c}^2}$. 
We choose to
perform an extrapolation of the result
in the whole kinematic region, obtaining
the  form factor  depicted in fig.2. The extrapolation provides a 
form factor having a nearly linear (with a small curvature term) 
$y-$dependence
in the kinematic range of the decays $B_c \to \eta_c \ell \nu$ and
 $B_c \to J/\psi \ell \nu$.

The same method and the same formulae can be used to calculate the form
factor $\Delta^\prime$ of $B_c \to \eta_c^\prime$ and $B_c \to \psi(2S)$; 
the only new ingredient is the wave-function
of the  $\psi(2S)$ radial excitation. Due to the
oscillating behaviour of $u_{\psi(2S)}$, the function  $\Delta^\prime$
is suppressed with respect to $\Delta$; 
interestingly enough, it has a negligible $y-$dependence, 
as one can observe in fig.2.
 
Before discussing the phenomenology of the decays
$B_c \to \eta_c (J/\psi) \ell \nu$ and 
$B_c \to \eta_c^\prime (\psi(2S)) \ell \nu$,
let us consider the matrix elements relevant for the transitions
$B_c \to B_s (B_s^*)$. 
A feature of the model we are considering is that both heavy-heavy and 
heavy-light mesons are described by the same formalism.
Therefore, eq.(\ref{delta}) can be  applied to calculate
$\Omega_1^s(\vec q =0)$, substituting $m_b$ with $m_s$ and the wave-function
$u_{\eta_c}$ with
$u_{B_s}$. In the limit $m_s \to 0$ and for a large value of the  
$b-$quark mass, eq.(\ref{delta}) becomes:
\begin{equation}
\Omega_1^s(\vec q=0) = {1 \over \sqrt 2} 
 {1 \over \sqrt { 2 M_{B_c} 2 M_{B_s}}} 
\int d \vec x ~~\Psi_{B_c} (\vec x) ~~\Psi^*_{B_s} (\vec x) \;\;\; ,
\label{omega1} \end{equation}
which differs by a factor ${1 \over \sqrt 2}$ with respect to the analogous
relation for $\Delta$. This factor is a 
consequence of considering a heavy-light 
meson in the final state instead of a heavy-heavy meson, 
and produces a suppression of the corresponding form factor.
Eq.(\ref{omega1}) suggests that, for similar (modulo the normalization 
condition) $B_c$ and $B_s$ wave-functions, the form factor 
$\Omega_1^s(\vec q=0)$ is close to the value $\Omega_1^s(\vec q=0)=1/\sqrt 2$.
The actual value, reported in Table \ref{t:tab0}, differs from this value
by a $7 \%$ effect.

The two results $\Delta(\vec q=0)\simeq 1$ and 
$\Omega_1^s(\vec q=0)\simeq 1/\sqrt 2$ are the main predictions of our 
analysis. They would deserve independent checks by different theoretical 
methods, namely by QCD sum rules in the  heavy quark limit.

From eq.(\ref{omega1}) it is also possible to derive a
relation, proposed in \cite{jenkins},
between the form factor
$\Omega_1^s$ and  the leptonic constant of the $B_s$ meson.
As a matter of fact, in the framework of the constituent 
quark model, the  $B_s$ 
leptonic constant,
defined by the matrix element: $<0|A_\mu|B_s(p)>=i f_{B_s} p_\mu$, is given by
\cite{pietroni}:
\begin{equation}
f_{B_s} = {\sqrt 3 \over 2 \pi M_{B_s}}
\int_0^\infty dk ~k ~u_{B_s} (k) 
\Big[{(E_b+ m_b)(E_s+ m_s)  \over E_b E_s }\Big]^{1/2}
[ 1 - {k^2 \over (E_b+ m_b)(E_s+ m_s)}] \;\;\;.
\label{fbs}
\end{equation}
For vanishing $m_s$ and  large $m_b$, $f_{B_s}$ is simply
related to the $B_s$ wave-function at the origin:
\begin{equation}
f_{B_s} = {\sqrt 3 \over M_{B_s}} \Psi_{B_s}(0) \;\;\; ,
\label{fbs1}
\end{equation}
a relation analogous to the van Royen-Weisskopf formula for the
quarkonium state.
Expanding $\Psi_{B_s}(x)$ near the origin in (\ref{omega1}),
 we obtain:
\begin{equation}
\Omega_1^s(\vec q=0) \simeq {1 \over 2 \sqrt 3} 
 f_{B_s} \sqrt M_{B_s} {1 \over \sqrt { 2 M_{B_c}}} 
\int d \vec x \;\; \Psi_{B_c} (\vec x) \;\; + \; corrections
\;\;\;. \label{omega1a}
\end{equation}
The numerical comparison of (\ref{omega1a}) with (\ref{omega1}), however,
suggests that the next-to-leading corrections in (\ref{omega1a})
 are sizeable, and therefore
the expansion (truncated at the first term) leading to eq.(\ref{omega1a})
appears to be of limited usefulness.

The value of $\Omega_1^s$ at zero-recoil is reported in Table
\ref{t:tab0}, and the plot of the form factor, extrapolated in the whole
kinematic region, is depicted in fig.2; the form factor presents a soft 
$y$-dependence in the narrow kinematic range spanned by the semileptonic
$B_c \to B_s, B_s^*$ transitions.

The same procedure can be applied to compute 
$\Omega_1^d$ and $\Sigma_1$, and the results are also depicted in
fig.2. The only new information is that, keeping finite values of the 
light quark masses,  a $SU(3)_F$ breaking effect 
between $\Omega_1^d$ and $\Omega_1^s$
of less than $3 \%$ is predicted. 

All the invariant functions
 can be  represented by the three-parameter formula
\begin{equation}
F(y)=F(0) \Big (1 - \rho^2 (y-1) + c \; (y-1)^2 \Big)
\label{fpar}
\end{equation}
in terms of the value at zero-recoil,  the slope $\rho^2$ and  the 
curvature $c$; the corresponding values
are collected in Table \ref{t:tab0}. 

A remark concerns the invariant functions $\Omega_2^{s,d}$ and $\Sigma_2$.
As mentioned in Sect.II, such form factors do not contribute at the 
zero-recoil point, since they appear in the term proportional to the 
small momentum $q$. In our approach,  based on considering
overlap integrals of wave-functions of mesons at rest, we cannot 
provide an independent calculation of
$\Omega_2^{s,d}$ and $\Sigma_2$, which therefore
will be neglected in our analysis. Such an approximation, however, 
could have relevant consequences only
in the case of the transitions $B_c \to D^{(*)} \ell \nu$; 
as already underlined, for the 
decays  $B_c \to B_s^{(*)}$ and $B_c \to B^{(*)}$ the
contribution from $\Omega_2$ is always proportional to the momentum $q$,
which  remains small in these processes.

Let us conclude the section comparing our
form factors $\Delta$, $\Omega_1^a$ and $\Sigma_1$ with the outcome of 
 the ISGW model \cite{isgw}, 
which has been widely applied to describe the 
heavy meson decays. In the ISGW approach,  the form factors 
 exponentially depend
on the squared momentum transfer to the lepton pair, and at zero-recoil they 
are given by products of parameters
relative to the mesons involved in the decays.
We depict in fig.2 the various invariant functions obtained in this
approach, observing
some agreement with our results in the case of  $\Delta$; 
as for $\Omega_1^s$, the result based on
\cite{isgw} deviates considerably from the value $1/\sqrt2$ suggested by 
our model.

\section{$B_c$ semileptonic decays}\label{s:semilep}

The form factors $\Omega_1^s$ and $\Omega_1^d$,
$\Delta$, $\Delta^\prime$and $\Sigma_1$
can be used to predict
the semileptonic $B_c$ decay
rates, as well as various decay distributions. 
Before doing  the calculation let us stress again that an
extrapolation is performed for the relevant matrix elements far from the
symmetry point (zero-recoil) where the form factors are originally
computed. Such a procedure would require the calculation of the
corrections, which could be sizable far from the symmetry point, an analysis
beyond the aim of the present work.
Considering the small range of momentum transfer
$t$ involved in $c \to
(s,d)$ transitions, it is plausible
that the extrapolation is quite under control for the
decays  $B_c \to B_s^{(*)}
\ell {\bar \nu}$, $B_d^{(*)} \ell {\bar \nu}$.
As for  $B_c \to \eta_c,~ J/\psi \ell {\bar \nu}$, the extrapolation is
done on a wider range of momentum transfer to the lepton pair.
However,  also in this case it is interesting
to make predictions and to compare them  with the  experimental
results. 
Notice that we only consider massless charged leptons in the
final state. 

Concerning the parameters needed in the  analysis, 
we use the experimental values of the masses of $\eta_c$, $J/\psi$, $\psi(2S)$,
$D^{(*)}$, $B^{(*)}$ and $B_s$ mesons; for the $\eta_c^\prime$ we use
$M_{\eta_c^\prime}=3.66$ GeV, and for $M_{B_s^*}$ we put:
$M_{B_s^*}=M_{B_s} + (M_{B^*_d} - M_{B_d})$.
For the CKM matrix elements we use 
$V_{cb}=0.039$ and
$V_{ub}=0.0032$; the values of $V_{cs}$ and $V_{cd}$ are fixed to
$V_{cs}=0.975$ and $V_{cd}=0.22$. The results for the decay widths 
are reported in Table \ref{t:tab1} where we also
report the corresponding branching fractions, obtained 
assuming for $\tau_{B_c}$ the CDF central
value: $\tau_{B_c}=0.46$ ps. 

In order to understand the effect of the $t-$dependence of the form factors, 
we also report in Table \ref{t:tab1} the results obtained assuming 
$t-$independent invariant functions, 
with the values fixed at the zero-recoil point.
The results provide us with an  upper bound for the various decay widths.
As expected, the momentum transfer dependence is mild in the case of the
$B_c \to B_s^{(*)}, B_d^{(*)}$ decays, 
where it only provides an effect of less than 
$10 \%$ in the decay rates. This is mainly due to the narrow $t-$ range spanned
in such decay modes. In the case of $B_c \to \eta_c$ and $J/\psi$, there is
a sizeable effect due to the $t-$ dependence of the form factors. On the
contrary, in the case of decays into radial 
excited states, $\eta_c^\prime$ and $\psi(2S)$, the $t$ dependence is
negligible. The $t-$dependence is important
for the Cabibbo suppressed  $B_c$ decays into $D$ and $D^*$.

From Table \ref{t:tab1} we conclude that
the semileptonic modes are dominated by two channels,
$B_c \to B_s \ell \nu$ and $B_c \to B_s^* \ell \nu$, in spite of the small 
phase space available for both the transitions; the two modes nearly represent
the $60 \%$ of the semileptonic width, a result in  agreement with
the predictions available in the literature.
 
As for the $b\to c$ induced semileptonic $B_c$ transitions, 
a peculiar role is played by the $B_c$ decay into $J/\psi$, due to the clear 
signature represented by three charged leptons from the same decay
vertex, two of them coming from $J/\psi$. This
signature has been exploited to identify the $B_c$ meson at Tevatron 
\cite{cdf}, and will be mainly
employed  at the future colliders \cite{sanchiz1}. 
Our prediction for the width of the decay $B_c \to J/\psi \ell \nu$ is:
$\Gamma(B_c \to J/\psi \ell \nu)\simeq 21.6 \times 10^{-15}$ GeV, with
an upper bound of $48 \times 10^{-15}$ GeV obtained using a $t-$independent
 form factor $\Delta$. The agreement of this result with other calculations 
in the literature suggests that the finite mass corrections, responsible
of subleading form factors in the matrix elements, should not be large. Tests
on the size of such corrections can be performed by measuring the $B_c$
decay rates  into longitudinally and transversely 
polarized $J/\psi$: $\Gamma_{L,T}=\Gamma(B_c \to J/\psi_{L,T} \ell \nu)$,
together with the corresponding decay distributions. Using the 
parameterization in (\ref{qqbar}) the decay widths are given by:
\begin{eqnarray}
\Gamma_L&=& {G_F^2 V_{cb}^2 M_{J/\psi}^5 \over 12 \pi^3} 
\int_1^{1+\delta} dy \; [\Delta(y)]^2 {\sqrt {y^2-1}} [r~y-1]^2
\nonumber \\
\Gamma_T&=& {G_F^2 V_{cb}^2 M_{J/\psi}^5 \over 12 \pi^3} 
\int_1^{1+\delta} dy \; [\Delta(y)]^2 {\sqrt {y^2-1}} [r^2+1-2~r~y]
\label{gaml}
\end{eqnarray}
where $r=M_{B_c}/M_{J/\psi}$ and 
$\delta={(M_{B_c}-M_{J/\psi})^2 \over 2 M_{B_c} M_{J/\psi}}$. 
The measurement of $d \Gamma_i/ dy$ provides information 
on  $\Delta$ and $V_{cb}$; in particular, 
if the  curvature term in $\Delta(y)$ is  neglected, the ratio
$\Gamma_T/\Gamma_L$ gives access to the slope $\rho^2$. 
The combination $V_{cb} \Delta(1)$ can be obtained  from the measurement
of $\Gamma_L$ and from the total width, and therefore a 
measurement of $V_{cb}$  is possible using this decay channel
\cite{sanchiz1,cinesi1}.
Such  new determinations of the CKM element 
$V_{cb}$, even though not accurate as from $B_d$ and $B_u$ decays,
would represent an important consistency check of the Standard Model.

Tests of the spin symmetry are provided by the measurement of the
decay distributions in the $y$ variable, whose deviations from the 
distributions related to a unique form factor $\Delta$ would imply the
presence of  spin symmetry-breaking terms. 

Let us finally observe that our prediction for the rates of the decays into
$0^-$ ($\bar c c$) states, 
$B_c \to \eta_c \ell \nu$ and $B_c \to \eta_c^\prime \ell \nu$, is smaller
than the value reported by other analyses.

\section{Non-leptonic $B_c$ decays}\label{s:nonlep}

Estimates of the decay rates of several two-body non-leptonic 
$B_c$ transitions can be obtained adopting
the factorization approximation. 
Such an approximation finds theoretical support 
in few  cases (large $N_c$ limit; $m_b \to \infty$ limit
in $ b \to u$ transitions involving heavy-light meson systems 
\cite{neubert1}); nevertheless, it
is widely used to estimate  non-leptonic decay rates
of mesons containing heavy quarks.

Let us first consider non-leptonic $B_c$ decay modes induced, at the quark
level, by the $b \to c$ and $u$ transitions. The effective Hamiltonian 
governing the processes reads:
\begin{eqnarray}
H_{eff}={G_F \over \sqrt{2}} &\big\{
&V_{cb}[c_1(\mu)Q_1^{cb}+c_2(\mu)Q_2^{cb}] 
+V_{ub}[c_1(\mu)Q_1^{ub}+c_2(\mu)Q_2^{ub}]+h.c. \big\} \nonumber \\
&+ & penguin \;\; operators \;\;\; ; \label{heff}
\end{eqnarray}
$G_F$ is the Fermi constant, $V_{ij}$ are CKM matrix
elements and $c_i(\mu)$ scale-dependent Wilson coefficients. The four-quark
operators $Q_1^{cb}$ and $Q_2^{cb}$ are given by
\begin{eqnarray}
Q_1^{cb}&=& \big[V_{ud}^* ~({\bar d}u)_{V-A}+V_{us}^* ~({\bar s}u)_{V-A}
+ V_{cd}^* ~({\bar d}c)_{V-A}+V_{cs}^*~ ({\bar s}c)_{V-A}\big]~ 
({\bar c} b)_{V-A} \nonumber \\
Q_2^{cb}&=& \big[V_{ud}^*~ ({\bar c}u)_{V-A} ~({\bar d}b)_{V-A}
+V_{us}^* ~({\bar c}u)_{V-A}~({\bar s}b)_{V-A}
+ V_{cd}^*~ ({\bar c}c)_{V-A}~({\bar d}b)
+V_{cs}^* ~({\bar c}c)_{V-A}~({\bar s}b)\big]  \nonumber \\
\label{q12} 
\end{eqnarray}
\noindent 
with $({\bar q}_1 q_2)_{V-A}={\bar q}_1 \gamma_\mu (1-\gamma_5)  q_2$;
analogous relations hold for $Q_1^{ub}$ and $Q_2^{ub}$.
 
As well known, 
the factorization approximation amounts to evaluate the matrix elements of
the four-quark operators in  (\ref{q12}) between the initial $B_c$
state and the final two-body hadronic states as the product of 
quark-current matrix elements. We adopt this approximation in the
calculation of the rates, neglecting the contribution of 
penguin operators, since their Wilson coefficients are small
with respect to $c_1$ and $c_2$ (interference effects of penguin diagrams
are of prime importance in producing $CP$ violating
asymmetries in $B_c$ decays). Moreover,
we do not take into account
 the weak annihilation contribution represented by a
$B_c$
meson annihilating into a charged $W$; in this amplitude, the final 
hadronic state is entirely produced out of the vacuum, and therefore the
contribution should be characterized by a sizeable form factor suppression. 
Annihilation processes  are  presumably relevant  mainly for 
rare or suppressed $B_c$ decays; in these cases  they 
deserve a dedicated analysis.

A further remark concerns the Wilson coefficients $c_1(\mu)$ and 
$c_2(\mu)$. Writing
the factorized amplitudes and taking into account the contribution of the
Fierz reordered currents, it turns out that the relevant coefficients are 
the combinations:
$a_1=c_1+ \xi c_2$ and $a_2=c_2+ \xi c_1$, with the QCD parameter $\xi$ given 
by
$\xi=1/N_c$. Several discussions concerning this parameter are available 
in the literature. 
We choose  $a_1=c_1$ and $a_2=c_2$, i.e. $\xi=0$,
in the spirit of the large $N_c$
limit, and use $c_1$ and $c_2$ computed at an energy scale of the
order of $m_b$. 
A detailed analysis of $1/N_c$ corrections 
to the coefficients $a_1, a_2$ as well as of the role of color-octet
current operators in $B$ decays
can be found in
\cite{neubertstech}. 
Analogous considerations hold 
for the decays induced by the $c \to s(d)$
transitions; in this case we choose the coefficients $c_1$ and $c_2$ at 
the scale of the charm mass.

The factorized amplitudes can be expressed in terms of 
the form factors in eqs.(\ref{ba}), (\ref{d}) and (\ref{qqbar}),
and of leptonic decay constants defined by the matrix 
elements: $<0|A_\mu|M(p)>=i f_M p_\mu$ and 
$<0|V_\mu|V(p,\epsilon)>=f_V M_V \epsilon_\mu$.
We use the values:
$f_{\pi^+}=0.131$ GeV,  
$f_{\rho^+}=0.208$ GeV and
$f_{a_1}=0.229$ GeV;
$f_{K^+}=0.159$ GeV,
$f_{K^{*+}}=0.214$ GeV and
$f_{K_1}=0.229$ GeV;
$f_{\eta_c}=0.31$ GeV,
$f_{\eta_c^\prime}=0.23$  GeV, 
$f_{\psi}=0.38$  GeV,
$f_{\psi^\prime}=0.28$ GeV, and finally
$f_{D}=0.2$ GeV,
$f_{D_s}=0.24$ GeV and
$f_{D^*}=0.23$ GeV,
$f_{D^*_s}=0.275$  GeV. 
Such values correspond to  experimental results or  to average values from
lattice QCD and QCD sum rules 
\footnote{A description of the current theoretical situation concerning 
the heavy meson leptonic decay constants is reported in 
the Appendices C and D of ref.\cite{babar}.}.

The decay rates of several non-leptonic $B_c$ transitions, obtained
using $c_1(m_b)=1.132$, $c_2(m_b)=-0.286$ and 
$c_1(m_c)=1.351$, $c_2(m_c)=-0.631$,
are collected in  Tables \ref{t:tab2}, \ref{t:tab3}. Also in this case
we use the physical phase space together with the expression of the matrix 
elements in (\ref{ba})-(\ref{qqbar}).

Few comments are in order. We observe 
the dominance of the decay modes induced by the charm transition, 
and in particular of the 
channel $B_c^+ \to B_s^* \rho^+$, which represents more than $10 \%$ of the
total $B_c$ width. It would be interesting to experimentally confirm
this prediction, even though the final state presents severe reconstruction 
difficulties. From the experimental point of view, more promising are 
the decay modes having a $J/\psi$ meson in the final state; among such modes, 
the decay channels $B_c^+ \to J/\psi \pi^+$  and
 $B_c^+ \to J/\psi \rho^+$ are particularly useful for the precise
measurement of the $B_c$ mass, by the
 complete reconstruction of the final state.
Also the decay into $a_1$ is of particular interest, due to the large decay 
rate.

Several
tests of factorization can be carried out, mainly using the decay channels 
having a $J/\psi$ in the final state.
For example, the assumption of the factorization approximation, 
together with the heavy quark spin symmetry, implies that the relation
\begin{equation}
{\Gamma(B_c^+ \to J/\psi \pi^+) \over 
{d\Gamma(B_c^+ \to J/\psi \ell^+ \nu)\over dy }|_{y=y_\pi}} = 
{3 \pi^2 V_{ud}^2 a_1^2 f_\pi^2 \over M_{B_c} M_{J/\psi} }
\label{fact1}
\end{equation}
holds in the limit $M_\pi \to 0$ 
($y_\pi={M_{B_c}^2+ M_{J/\psi}^2 \over 2 M_{B_c} M_{J/\psi}}$).
An analogous relation
holds for the $B_c$ transition into the radial excited state $\psi(2S)$:
\begin{equation}
{\Gamma(B_c^+ \to \psi(2S) \pi^+) \over 
{d\Gamma(B_c^+ \to \psi(2S) \ell^+ \nu)\over dy}|_{y=y_\pi}} = 
{3 \pi^2 V_{ud}^2 a_1^2 f_\pi^2 \over M_{B_c} M_{\psi(2S)} } \;\;\; .
\label{fact2}
\end{equation} 

In the case of a $\rho$ meson in the final state one has:
\begin{eqnarray}
{\Gamma(B_c^+ \to J/\psi \rho^+) \over 
{d\Gamma(B_c^+ \to J/\psi \ell^+ \nu)\over dy}|_{y=y_\rho} } &=& 
{3 \pi^2 V_{ud}^2 a_1^2 f_\rho^2 [8 M_{J/\Psi}^2 M_\rho^2 +  
(M_{B_c}^2 - M_{J/\psi}^2 -M_\rho^2)^2 ] \over 8 M_{B_c}^2 M_{J/\psi}^5 } 
\nonumber \\
&\times& {\lambda^{1 \over 2}(M_{B_c}^2,M_{J/\psi}^2,M_\rho^2) \over 
\sqrt{y^2-1} [r^2 y_\rho^2 - 6 r y_\rho + 2 r^2 +3]} \;\;\; ,
\label{fact3}
\end{eqnarray}
$\lambda$ being the triangular function, 
$r={M_{B_c} \over M_{J/\psi}}$ and
$y_\rho={M_{B_c}^2+ M_{J/\psi}^2 -M_\rho^2\over 2 M_{B_c} M_{J/\psi}}$.

To test eqs.(\ref{fact1})-(\ref{fact3})  two-body
decay rates and  the differential $B_c^+ \to J/\psi \ell^+ \nu$ decay width
are required; the measurement of such quantities, possible at 
 the hadronic facilities, would provide us with
important information 
on the heavy quark
spin symmetry as well as on the factorization approximation in $B_c$ 
decays.

\section{Conclusions}

We have presented a determination 
of the invariant functions parameterizing the semileptonic $B_c$ matrix 
elements in the infinite heavy quark mass limit. The form factors are obtained
as overlap integrals of meson wave-functions, obtained in the framework of a 
QCD relativistic potential model. An interesting result is that, although 
not constrained by symmetry arguments, the normalization of the form factor
$\Delta$ describing the transition  $B_c\to J/\psi \ell \nu$ is close to 1
at the zero-recoil point, as being the overlap of similar wave-functions. 
On the contrary, the form factors relative to
 the transitions into heavy-light mesons, at zero-recoil point,
are suppressed by a factor $\simeq 1/\sqrt2$ with respect to $\Delta$. 
These results have several phenomenological consequences, 
in semileptonic and  non-leptonic $B_c$ decay processes, which can be
experimentally tested.
Moreover, they affect other important processes, 
such as  radiative flavour-changing $B_c$ decays \cite{fajfer} and 
CP violating $B_c$ transitions \cite{masetti,liu}. In particular, the 
invariant functions computed in this paper can be useful to identify
the $B_c$ decay channels characterized by a clean experimental signature, a
large branching fraction and a visible CP asymmetry; the identification of
this kind of decay modes is of 
paramount importance for the physics program of the experiments at the future
accelerators.

\vspace*{1cm}
\noindent {\bf Acknowledgments\\}
\noindent 
(F.D.F.) thanks Prof. R. Gatto for hospitality at 
D\'epartement de Physique Th\'eorique,
Universit\'e de Gen\`eve,  and for interesting discussions and
encouragement. She also acknowledges "Fondazione Angelo Della Riccia"
for a fellowship. 

\clearpage
\begin{table}
\caption{ Parameters of the form factors ($\psi'=\psi(2S)$). 
The functional dependence is  in 
(\ref{fpar}).\\}
\vspace*{0.5cm}
\begin{tabular}{|l|c|c|c|c|}
~~~Channel                & Form factor~~~ &
$~F(1)~~~~~$ & $~\rho^2~~~~$ &$~c~~~~$\\ \tableline
$B_c \to B_s (B_s^*)$     & $\Omega_1^s$& 
$0.66$       & $8$          &$0$       \\
$B_c \to B_d (B_d^*)$     & $\Omega_1^d$&
$0.66$       & $8$          &$0$       \\
$B_c \to \eta_c (J/\psi)$ & $\Delta$&
$0.94$       & $2.9$           &$3$       \\
$B_c \to \eta_c^\prime (\psi^\prime)$&$\Delta^\prime$&
$0.23$&$0$         &$0$       \\
$B_c \to D (D^*)$&$\Sigma_1$&
          $0.59$       &$1.3$          &$0.4$
\end{tabular}
\label{t:tab0}
\end{table}
\vskip 2cm
\begin{table}
\caption{Semileptonic $B_c^+$ decay widths and branching fractions.\\}
\vspace*{0.5cm}
\begin{tabular}{|l|c|c|c|c|} 
~~Channel &$\Gamma   (10^{-15}$ GeV) &$\Gamma_L (10^{-15}$ GeV)
          &$\Gamma_T (10^{-15}$ GeV) &BR \\ \tableline
$B_c^+ \to B_s e^+ \nu  $&$11.1 (12.9) $&-&-&$0.8 (0.9)\times 10^{-2}$\\
$B_c^+ \to B^*_s e^+ \nu$&$33.5 (37.0) $&$19.1 (21.4)$&$7.2 (7.8)$&$2.3
(2.5) \times 10^{-2}$
\\ \tableline
$B_c^+ \to B_d e^+ \nu  $&$0.9 (1.0) $&-&-&$0.06 (0.07) \times 10^{-2}$\\
$B_c^+ \to B^*_d e^+ \nu$&$2.8 (3.2) $&$1.6 (1.8)$&$0.6 (0.8)$&$0.19 (0.22)
 \times 10^{-2}$ \\ 
\tableline
$B_c^+ \to \eta_c e^+ \nu  $&$2.1 (6.9)  $&-&-&$0.15 (0.5)\times 10^{-2}$\\
$B_c^+ \to J/\psi e^+ \nu$ &$21.6 (48.3) $&$13.2 (33.2)$&$4.2 (7.6)$&$1.5
(3.3) \times 10^{-2}$\\
\tableline
$B_c^+ \to \eta_c^\prime e^+ \nu  $&$0.3 (0.3)  $&-&-&$0.02 (0.02)\times
10^{-2}$\\
$B_c^+ \to \psi^\prime e^+ \nu$ &$1.7 (1.7)  $&$1.1 (1.1)$&$0.3
(0.3)$&$0.12
(0.12) \times 10^{-2}$\\
\tableline
$B_c^+ \to D^0 e^+ \nu   $&$ 0.005 (0.03) $&-&-&$0.0003 (0.002) \times
10^{-2}$\\
$B_c^+ \to D^{*0} e^+ \nu$&$0.12 (0.5) $&$0.08 (0.35)$&$0.02 (0.05)$&$0.008
(0.03) \times 10^{-2}$ 
\end{tabular}
\label{t:tab1}
\end{table}

\clearpage

\baselineskip 18pt
\begin{table}
\caption{Non-leptonic ($b \to c,u$) $B_c^+$ decay widths and branching 
fractions.\\}
\vspace*{0.5cm}
\begin{tabular}{|c|c|c||c|c|c|}
~~Channel~~ &$\Gamma   (10^{-15}$ GeV) &$BR $ & 
~~Channel~~ &$\Gamma   (10^{-15}$ GeV) &$BR $ \\
\hline
$ \eta_c \pi^+  $&$ a_1^2 ~ 0.28 $& $2.6 \times 10^{-4}$
&$\eta_c K^+$& $a_1^2~ 0.023$&$2 \times 10^{-5}$\\
$ \eta_c \rho^+  $&$a_1^2 ~ 0.75 $& $6.7 \times 10^{-4}$
&$\eta_c K^{*+}$& $a_1^2~ 0.041$&$3.6 \times 10^{-5}$\\
$ \eta_c a_1^+  $&$a_1^2 ~ 0.96 $& $8.6 \times 10^{-4}$
&$\eta_c K_1^+$& $a_1^2 ~0.05$&$4.4 \times 10^{-5}$\\
&&&&&\\
$ \eta_c^\prime \pi^+  $&$a_1^2 ~ 0.074 $&$6.6 \times 10^{-5}$
 &$\eta_c^\prime K^+$& $a_1^2~
0.0055$&$5 \times 10^{-6}$\\
$ \eta_c^\prime \rho^+  $&$a_1^2 ~ 0.16 $& $1.5 \times 10^{-4}$
&$\eta_c^\prime K^{*+}$& $a_1^2
~0.008$&$7.4 \times 10^{-6}$\\
$ \eta_c^\prime a_1^+  $&$a_1^2 ~ 0.15 $& $1.4 \times 10^{-4}$
&$\eta_c^\prime K_1^+$& $a_1^2
~0.0075$&$6.7 \times 10^{-6}$\\
&&&&&\\
$ J/\psi \pi^+  $&$a_1^2 ~ 1.48 $& $1.3 \times 10^{-3}$
&$J/\psi K^+$& $a_1^2 ~0.076$&$6.8 \times 10^{-5}$\\
$  J/\psi \rho^+  $&$a_1^2 ~ 4.14 $& $3.7 \times 10^{-3}$
&$J/\psi K^{*+}$& $a_1^2~ 0.23$&$2 \times 10^{-4}$\\
$ J/\psi a_1^+  $&$a_1^2 ~ 5.78 $&$5.2 \times 10^{-3}$
 &$J/\psi K_1^+$& $a_1^2 ~0.3 $&$2.7 \times 10^{-4}$\\
&&&&&\\
$ \psi^\prime \pi^+  $&$a_1^2 ~ 0.22 $& $1.9 \times 10^{-4}$
&$\psi^\prime K^+$& $a_1^2 ~0.01$
&$9.3 \times 10^{-6}$\\
$ \psi^\prime \rho^+  $&$a_1^2 ~ 0.54 $&$4.86 \times 10^{-4}$
&$\psi^\prime
K^{*+}$& $a_1^2~
0.03$&$2.6 \times 10^{-5}$\\
$ \psi^\prime a_1^+  $&$a_1^2 ~ 0.65 $& $5.8 \times 10^{-4}$
&$\psi^\prime K_1^+$& $a_1^2~
0.033$&$3 \times 10^{-5}$\\
&&&&&\\
$D^+ {\bar D}^0$ & $a_2^2 ~ 0.15 $& $8.4 \times 10^{-6}$
& $D^+_s {\bar D}^0$ & $a_2^2 ~ 0.01
$&$6 \times 10^{-7}$\\
$D^+ {\bar D}^{*0}$ & $a_2^2 ~ 0.13 $& $7.5 \times 10^{-6}$
& $D^+_s {\bar D}^{*0}$ & $a_2^2 ~
0.009 $&$5.3 \times 10^{-7}$\\
$D^{*+} {\bar D}^0$ & $a_2^2 ~ 1.46 $&$8.4 \times 10^{-5}$ 
& $D^{*+}_s {\bar D}^0$ & $a_2^2 ~
0.087$&$5 \times 10^{-6}$\\
$D^{*+} {\bar D}^{*0}$ & $a_2^2 ~ 2.4 $& $51.4 \times 10^{-4}$
& $D^{*+}_s {\bar D}^{*0}$ &
$a_2^2 ~0.15 $&$8.4 \times 10^{-6}$\\
&&&&&\\
$\eta_c D_s $ & $ (a_1 ~ 7.8+ a_2 ~ 1.6)^2 \times 10^{-1}$ &$5 \times
10^{-3}$ &
$\eta_c D^+ $ & $ (a_1 ~ 0.86+ a_2 ~ 0.46)^2 \times 10^{-1}$ &$
5 \times 10^{-5}$ \\
$\eta_c D_s^* $ & $ (a_1 ~ 3.6+a_2 ~ 6.05)^2\times 10^{-1}$ & 
$3.8 \times 10^{-4}$&
$\eta_c D^{*+} $ & $ (a_1 ~ 0.7+ a_2 ~ 0.9)^2 \times 10^{-1}$ &
$2 \times 10^{-5}$ \\
$\eta_c^\prime D_s $ & $ (a_1 ~ 1.5+a_2~ 3.2)^2\times 10^{-1}$ &
$3.7 \times 10^{-5}$ &
$\eta_c^\prime D^+ $ & $ (a_1 ~ 0.28+ a_2 ~ 0.7)^2 \times 10^{-1}$ & 
$1 \times 10^{-6}$\\
$\eta_c^\prime D_s^* $ & $ (a_1 ~ 0.79+a_2~1.8)^2 \times 10^{-1}$ &
$1 \times 10^{-5}$ &
$\eta_c^\prime D^{*+} $ & $ (a_1 ~ 0.17+ a_2 ~ 0.8)^2 \times 10^{-1}$ &$6 
\times 10^{-8}$
\\
&&&&&\\
$J/\psi D_s $ & $ (a_1 ~ 6.7+a_2~2.3)^2\times 10^{-1}$ &$3.4 \times
10^{-3}$
&
$J/\psi D^+ $ & $ (a_1 ~ 1.31+a_2~0.47)^2\times 10^{-1}$ &$1.3 \times
10^{-4}$ \\
$J/\psi D_s^* $ & $ (a_1 ~ 11+a_2~ 10.4)^2\times 10^{-1}$ & $5.9 \times
10^{-3}$&
$J/\psi D^{*+} $ & $ (a_1 ~ 2.02+a_2~2.3)^2\times 10^{-1}$ & $1.9 \times
10^{-4}$\\
$\psi^\prime D_s $ & $ (a_1 ~ 1.4+a_2~ 1.33)^2\times 10^{-1}$ &$1 
\times10^{-4}$ &
$\psi^\prime D^+ $ & $ (a_1 ~ 0.35+a_2~0.36)^2\times 10^{-1}$ &
$5.8 \times 10^{-6}$ \\
$\psi^\prime D_s^* $ & $ (a_1 ~ 2.75+a_2~ 7.8)^2\times 10^{-1}$ &$5.7
\times 10^{-5}$ &
$\psi^\prime D^{*+} $ & $ (a_1 ~ 0.55+a_2~1.76)^2\times 10^{-1}$ & 
$8.7 \times 10^{-7}$
\end{tabular}
\label{t:tab2}
\end{table}

\clearpage

\begin{table}
\caption{Non-leptonic ($c \to s,d$) $B_c^+$ decay widths and branching 
fractions.\\}
\vspace*{0.5 cm}
\begin{tabular}{|c|c|c||c|c|c|}
~~Channel~~&$\Gamma   (10^{-15}$ GeV)~ &$BR$&
~~Channel~~&$\Gamma   (10^{-15}$ GeV)~ &$BR$ \\ \tableline
$B_s \pi^+$ & $a_1^2 ~ 30.6$ &$4 \times 10^{-2}$~~ & 
$B_s K^+$   & $a_1^2 ~ 2.15$  &$2.7 \times 10^{-3}$~~ \\ 
$B_s \rho^+$& $a_1^2 ~ 13.6$ &$1.7 \times 10^{-2}$~~ & 
$B_s K^{*+}$& $a_1^2 ~ 0.043$&$5.4 \times 10^{-5}$~~\\ 
$B_s^* \pi^+$&$a_1^2 ~ 35.6$ &$4.5 \times 10^{-2}$~~ & 
$B_s^* K^+$&  $a_1^2 ~ 1.6$  &$2 \times 10^{-3}$~~\\ 
$B_s^* \rho^+$ &$a_1^2 ~ 110.1$&$1.4 \times 10^{-1}$& & &\\ 
&&&&&\\
$B_d \pi^+$ & $a_1^2 ~ 1.97$ &  $2.5 \times 10^{-3}$~~ & 
$B_d K^+$   & $a_1^2 ~ 0.14$ &  $1.8 \times 10^{-4}$~~ \\ 
$B_d \rho^+$& $a_1^2 ~ 1.54$ &  $2 \times 10^{-3}$~~ & 
$B_d K^{*+}$ &$a_1^2 ~ 0.032$ &  $4 \times 10^{-5}$~~ \\ 
$B_d^* \pi^+$&$a_1^2 ~ 2.4$  &  $3 \times 10^{-3}$~~ & 
$B_d^* K^+$  &$a_1^2 ~ 0.12$ &  $1.6 \times 10^{-4}$~~ \\ 
$B_d^* \rho^+$&$a_1^2 ~ 8.6$ & $1 \times 10^{-2}$~~ & 
$B_d^* K^{*+}$ & $a_1^2 ~ 0.34$&$4.4 \times 10^{-4}$~~
\end{tabular}
\label{t:tab3}
\end{table}

\clearpage
\hskip 3 cm {\bf FIGURE CAPTIONS}
\vskip 1 cm
\noindent {\bf Fig. 1} \par
\noindent
Reduced $L=0$ wave-functions $u_M(k)$ of heavy-heavy 
($B_c$, $J/\Psi$, $\psi(2S)$)
and heavy-light ($B_s$, $B_d$, $D$)
mesons. The wave-functions are obtained by
solving the wave equation (\ref{7}); they
describe both the pseudoscalar $0^-$ and vector $1^-$ mesons.

\vskip 1 cm
\noindent {\bf Fig. 2} \par
\noindent
Form factors of $B_c$ semileptonic decays. 
The variable $y$ is related to the squared momentum $t$, transferred to the
lepton pair, by the relation:
$y={M_{B_c}^2 + M_{M}^2 -t \over 2 M_{B_c} M_{M} }$.
The solid lines correspond to the form factors 
obtained by the model discussed in the paper;
the dashed lines refer to the model in ref.\cite{isgw}.

\clearpage
\begin{figure}
\begin{center}
\epsfig{file=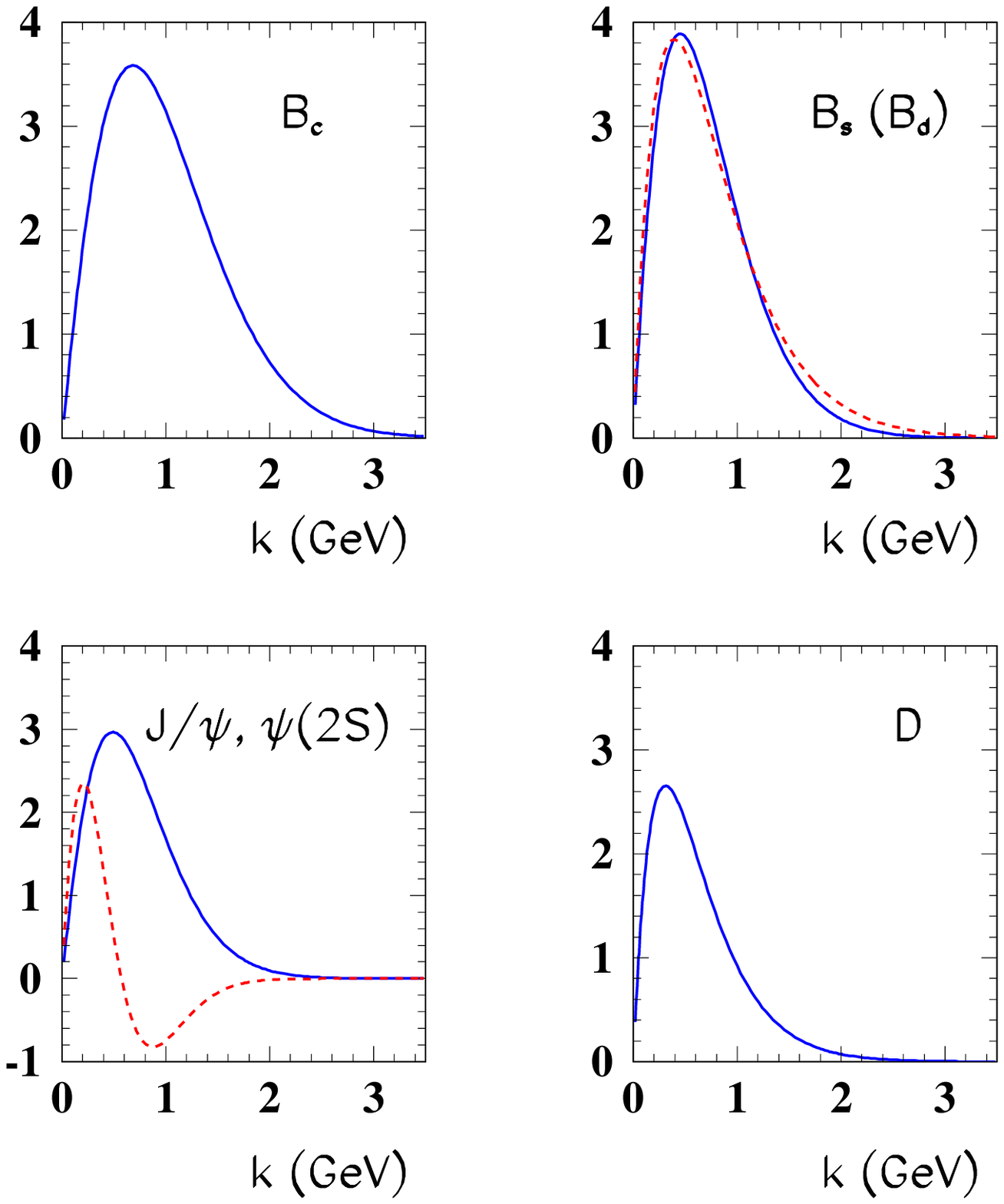,height=17cm} 
\label{f:wf}
\caption{}
\end{center}
\end{figure}
\vskip 1cm
\clearpage

\vskip 2cm
\begin{figure}
\begin{center}
\epsfig{file=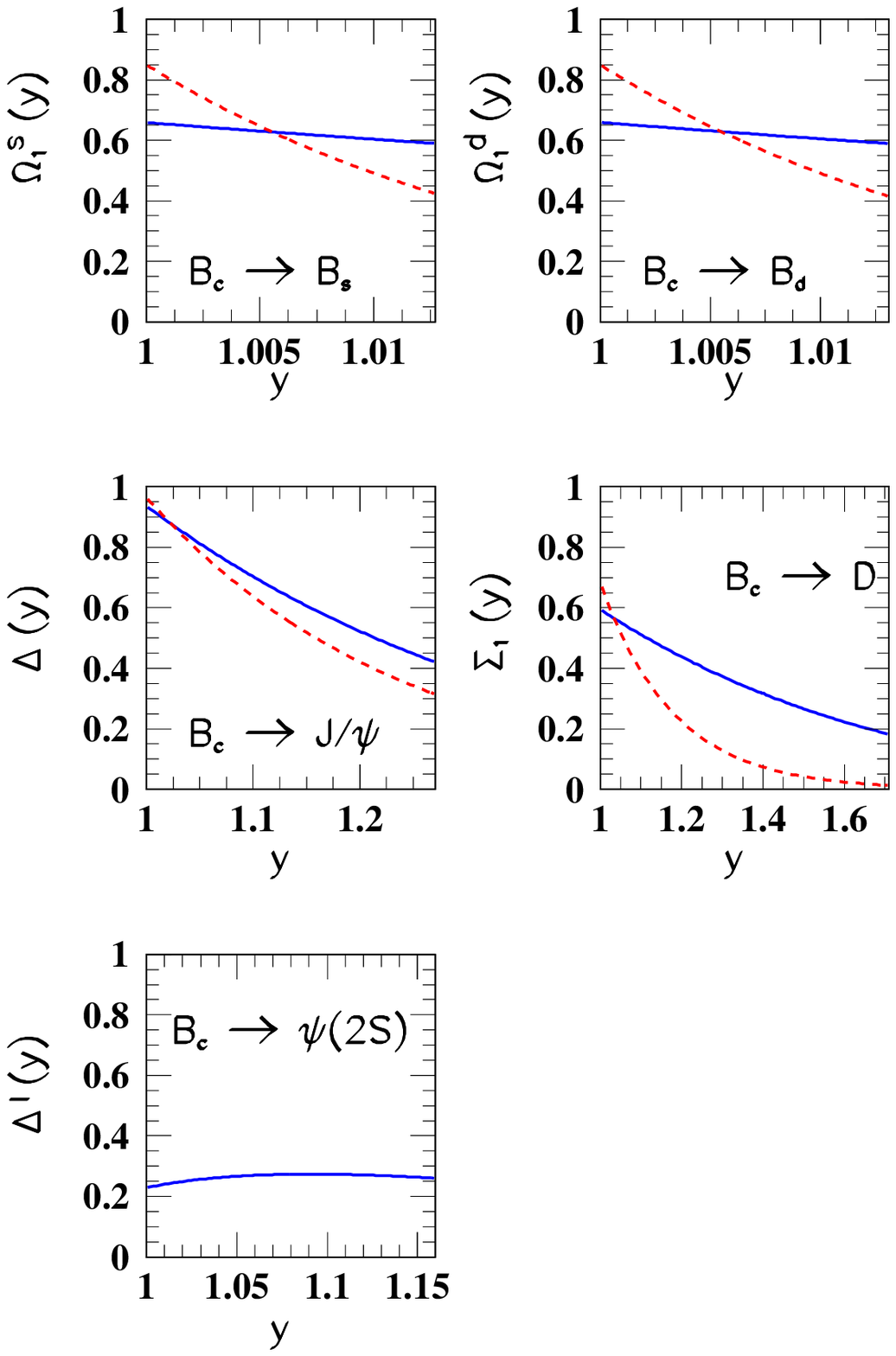,height=17cm}
\caption{}
\label{f:ff}
\end{center}
\end{figure}

\clearpage
\begin{references}
%
\bibitem{cdf}
CDF Collaboration, F. Abe {\it et al.}, Phys. Rev. Lett. {\bf 81}, 2432
(1998); Phys. Rev. D {\bf 58}, 112004 (1998).

\bibitem{bcprod}
K. Cheung, Phys. Rev. Lett. {\bf 71}, 3413 (1993);
K. Kolodziej, A. Leike and R. R\"uckl, Phys. Lett. B {\bf 355}, 337
(1995);
C.H. Chang, Y.Q. Chen, G.P. Han and H.T. Jiang, Phys. Lett. B {\bf 364}, 
78 (1995);
M. Masetti and F. Sartogo, Phys. Lett. B {\bf 357}, 659 (1995); 
C.H. Chang, Y.Q. Chen, and R.J. Oakes, Phys. Rev. D {\bf 54}, 4344 (1996);
A.V. Berezhnoy, V.V. Kiselev, A.K. Likhoded and A.I. Onishchenko, 
Yad. Fiz. {\bf 60}, 1866 (1997).

\bibitem{revbc}
S. Gershtein et al., hep-ph/9803433.

\bibitem{lus}
M. Lusignoli and M. Masetti, Z. Phys. C {\bf 51}, 549 (1991).

\bibitem{quigg}
C. Quigg, in {\it Proceedings of the Workshop on B Physics at Hadron
Accelerators}, Snowmass, Colorado 1993, edited by P. McBride and C.S.
Mishra, pag.439.

\bibitem{paver}
P. Colangelo, G. Nardulli and N. Paver, Z. Phys. C {\bf 57}, 43 (1993).

\bibitem{beneke}
M. Beneke and G. Buchalla,  Phys. Rev. D {\bf 53}, 4991 (1996).

\bibitem{ani}
A. Y. Anisimov, I.M. Narodetskii, C. Semay, B. Silvestre-Brac, Phys. Lett. B
{\bf 452}, 129 (1999).

\bibitem{woloshyn}
B.D. Jones and R.M. Woloshyn, Phys. Rev. D {\bf 60}, 014502 (1999).

\bibitem{svz}
M.A. Shifman, A.I. Vainshtein and V.I. Zakharov, Nucl. Phys. B {\bf 147}, 
385, 448 (1979); V.A. Novikov {\it et al.,}, Phys. Rept. {\bf 41}, 1 (1978).
 
\bibitem{oldbc}
L.J.Reinders, H.R.Rubinstein and S. Yazaki, Phys. Rep. {\bf 127},1 (1985);
V.V. Kiselev and A. Tkabladze, Sov. J. Nucl. Phys. {\bf 50} (6), 1063 (1989);
T.M. Aliev and O. Yilmaz, Nuovo Cimento A {\bf 105}, 827 (1992); 
M. Chabab, Phys. Lett. B {\bf 325}, 205 (1994).

\bibitem{bagan}
E. Bagan {\it et al.}, Zeit. Phys. C {\bf 64}, 57 (1994);
V.V. Kiselev and A. Tkabladze, Phys. Rev. D {\bf 48}, 5208 (1993).

\bibitem{kiselev}
V.V. Kiselev, A.K. Likhoded and A.I. Onishchenko, hep-ph/9905359.

\bibitem{bsw}
M. Wirbel, B. Stech and M. Bauer, Zeit. Phys. C {\bf 29}, 637 (1985).

\bibitem{isgw}
N. Isgur, D. Scora, B. Grinstein and M. Wise, Phys. Rev. D {\bf 39}, 799
(1989).

\bibitem{du}
D. Du and Z. Wang, Phys. Rev. D {\bf 39}, 1342 (1989).

\bibitem{chang} 
C.H. Chang and Y.Q. Chen, Phys. Rev. D {\bf 49}, 3399 (1994).

\bibitem{liu}
J.F. Liu and K.T. Chao, Phys. Rev. D {\bf 56}, 4133 (1997).

\bibitem{jenkins}
E. Jenkins, M. Luke, A. V. Manohar and M. Savage, 
Nucl. Phys. B {\bf 390}, 463 (1993). 

\bibitem{sanchiz}
M.A. Sanchiz-Lozano, Nucl. Phys. B {\bf 440}, 251 (1995).

\bibitem{pietroni}
P. Cea, P. Colangelo, L. Cosmai and G. Nardulli, Phys. Lett. B {\bf 206},
691 (1988);
P. Colangelo, G. Nardulli and M. Pietroni, Phys. Rev. D {\bf  43}, 3002
(1991).

\bibitem{trace}
A. F. Falk, H. Georgi, B. Grinstein and M.B. Wise, Nucl. Phys. B {\bf 
343}, 1 (1990); J.D. Bjorken, talk given at Les Rencontres de la Vall\'ee
d'Aoste, La Thuile, Italy, March 1990, SLAC-PUB-5278 (1990).

\bibitem{neu}
M. Neubert, Phys. Rept. {\bf  245}, 259 (1994).
 
\bibitem{cas}
R. Casalbuoni et al., Phys. Lett. B {\bf 303}, 95 (1993).

\bibitem{richardson}
J. L. Richardson, Phys. Lett. B {\bf 82}, 272 (1979).

\bibitem{tedesco}
P. Colangelo, G. Nardulli and L. Tedesco, Phys. Lett. B {\bf 272}, 344 (1991).

\bibitem{defazio}
P. Colangelo, F. De Fazio and G. Nardulli, Phys. Lett. B {\bf 334}, 
175 (1994); F. De Fazio, Mod. Phys. Lett. A {\bf 11}, 2693 (1996).

\bibitem{fdf}
P. Colangelo and F. De Fazio, hep-ph/9904363. 

\bibitem{qmodels}
W. Kwong and J. Rosner, Phys. Rev. D {\bf 44}, 212 (1991);
E. Eichten and C. Quigg, Phys. Rev. D {\bf 49}, 5845 (1994);
S.S. Gershtein et al., Phys. Rev. D {\bf  51}, 3613 (1995); 
L.P. Fulcher, Phys. Rev. D {\bf 60}, 074006 (1999), and references therein.

\bibitem{shanahan}
H.P. Shanahan, P.Boyle, C.T.H. Davies and H. Newton, UKQCD Collaboration, 
Phys. Lett. B {\bf 453}, 289 (1999).

\bibitem{francesi}
A. Le Yaouanc, L. Oliver, O. Pene and J.C. Raynal, 
{\it Hadron Transitions in the Quark Model}, Gordon and Breach, NY, (1988).

\bibitem{cinesi1}
M.T. Choi and J.K. Kim, Phys. Lett. B {\bf 419}, 377 (1998).

\bibitem{babar}
The BaBar Physics Book, 
P.F. Harrison and H. R. Quinn eds., SLAC-R-504 (October 1998).

\bibitem{sanchiz1}
M. Galdon and M.A. Sanchiz-Lozano, Zeit. Phys. C {\bf 71}, 277 (1996).

\bibitem{neubert1}
M. Beneke, G. Buchalla, M. Neubert and C.T. Sachraida, 
Phys. Rev. Lett. {\bf 83}, 1914 (1999).
 
\bibitem{neubertstech}
M. Neubert and B. Stech, in {\it Heavy Flavours II}, edited by A. J. Buras
and M. Lindner, World Scientific, Singapore, pag. 294.

\bibitem{fajfer}
D. Du, X. Li and Y. Yang, Phys. Lett. B {\bf 380}, 193 (1996);
S. Fajfer, D. Prelovsek and P. Singer, Phys. Rev. D {\bf 59}, 114003 (1999);
T.M. Aliev and M. Savci, hep-ph/9908203.

\bibitem{masetti}
M. Masetti, Phys. Lett. B {\bf 286}, 160 (1992);
Y.S. Dai and D.S. Du, Eur. Phys. J. C {\bf 9}, 557 (1999).



\end{references}
\end{document}